\documentclass[prl,showpacs,twocolumn,superscriptaddress,floatfix]{revtex4}

\usepackage{graphicx}
\usepackage{latexsym}

\begin{document}


\title{Spatial variability enhances species fitness in stochastic 
       predator--prey interactions}

\author{Ulrich Dobramysl} \email{ulrich.dobramysl@jku.at}
\affiliation{Institute of Semiconductor and Solid State Physics, 
  Johannes Kepler University, Altenbergerstr. 69, 4040 Linz, Austria}

\author{Uwe C. T\"auber} \email{tauber@vt.edu}
\affiliation{Department of Physics, Virginia Tech, Blacksburg, VA 24061-0435}

\begin{abstract}
We study the influence of spatially varying reaction rates on a spatial
stochastic two-species Lotka--Volterra lattice model for predator--prey 
interactions using two-dimensional Monte Carlo simulations. 
The effects of this quenched randomness on population densities, transient 
oscillations, spatial correlations, and invasion fronts are investigated. 
We find that spatial variability in the predation rate results in more
localized activity patches, which in turn causes a remarkable increase in the 
asymptotic population densities of both predators and prey, and accelerated 
front propagation.
\end{abstract}


\date{\today}

\maketitle


Understanding biological diversity has been a central issue in ecology 
\cite{May73, Maynard74, Sigmund98, Murray02}.
In order to understand the coexistence of competing species, several simplified
`toy' models for the dynamics of few interacting populations such as the 
paradigmatic Lotka--Volterra predator-prey model have been investigated.
More recently, the crucial role of spatial fluctuations and stochasticity in 
stabilizing such systems has been recognized \cite{Durrett99}.
Indeed, stochastic predator--prey models \cite{Matsuda92, Tome94, Boccara94} 
that consistently account for the internal reaction noise yet reduce to the 
classical coupled Lotka--Volterra differential equations in the well-mixed 
mean-field limit have been found to display a remarkable wealth of intriguing 
features \cite{Mobilia07}:
In contrast to the regular nonlinear oscillations of the deterministic 
Lotka--Volterra model which always entail a return to the initial state, these
stochastic spatial models yield long-lived, but ultimately decaying erratic 
population oscillations 
\cite{Provata99, Albano99, Lipowski99, Droz01, Antal01, Kowalik02}.
In the absence of spatial degrees of freedom, these oscillations can be 
understood through a resonant amplification mechanism for stochastic 
fluctuations that drastically extends the transient period before the (finite) 
system finally reaches the absorbing stationary state (predator extinction)
\cite{McKane05}.
In spatially extended systems, the mean-field Lotka--Volterra 
reaction-diffusion model is known to support traveling wave solutions 
\cite{Dunbar83, Sherratt97, Aguiar04}.
In corresponding stochastic spatial population models, spreading activity 
fronts induce persistent correlations between the prey and predator species,
and further enhance the amplitude and life time of local population 
oscillations \cite{Mobilia07, Washenberger07, GlobalOsc}.

In our studies of different stochastic spatial model variants for competing 
predator--prey populations, we have found these intriguing spatio-temporal 
structures and the overall features to be remarkably generic and robust with 
respect to even rather drastic changes of the detailed microscopic interaction 
rules \cite{Mobilia06, Washenberger07}.
Yet to render these models more realistic and relevant for biological systems,
one must obviously allow for different fitness of the individuals as well as
spatial variations in the rates that describe the population kinetics.
In this letter, we address the latter situation by considering the reaction 
rates to be quenched random variables, drawn from truncated Gaussian 
distributions.
This model can be interpreted as describing a direct environmental influence on
the species death and reproduction rates such as, e.g., a local variability of 
available resources.

By means of individual-based stochastic cellular automaton Monte Carlo
simulations we find that an increasing spatial variation of the predation 
interaction or species invasion rate (with fixed mean) enhances the 
steady-state population densities (which we take as a measure of the species' 
fitness) of {\em both} predators and prey.
In contrast, mere variations of the predator death and prey reproduction rates 
have very little effect.
While a simple mean-field averaging over varying predation rates does indeed 
predict a marked stationary density increase, it also grossly overestimates 
cooperative behavior and cannot adequately describe our numerical results.
In fact, we shall argue that the principal fitness enhancement mechanism rests
in the fact that stronger disorder in the predation rate reduces the size of 
the localized regions populated by both species, thus amplifying the initial 
local population oscillations and permitting a larger number of activity 
patches in the asymptotic long-time limit.
Thus, the fitness enhancement of both species through spatial variability, 
notably in the absence of any evolutionary adaption processes, is a consequence
of the emerging dynamical correlations.
Remarkably, we find that quenched randomness in the predation rates also
slightly increases the speed of spreading activity fronts.


We consider a stochastic Lotka--Volterra model on a square lattice (typically 
with 512 $\times$ 512 sites) with periodic boundary conditions.
Individuals of both particle species perform random walks through unbiased 
nearest-neighbor hopping (which occurs with probability one, so in effect all
rates listed below are to be understood as relative to the diffusivity $D$).
We allow multiple, essentially unrestricted lattice site occupation for 
particles of either or both species (the maximum number per site $i$ is capped 
at $n_i \leq 1000$).
This eliminates the predator extinction transition present in model variants 
with restricted site occupation \cite{Antal01, Mobilia07, Washenberger07}.
The `predator' species is subject to spontaneous decay $A \to \emptyset$ with
rate $\mu$, in contrast with the `prey' particles that may produce offspring 
$B \to 2 B$ with rate $\sigma$.
When individuals of both species meet on any lattice site, a prey is `eaten'
and the predators simultaneously reproduce, i.e., we implement the predation 
interaction $A + B \to 2 A$ with rate $\lambda$.
Our dynamical Monte Carlo simulation proceeds with random sequential updates; 
a Monte Carlo step (MCS) is completed once on average each particle in the 
system has been moved and had the chance to react \cite{Algorithm}.

Spatial variability is introduced by drawing the reaction probabilities for 
each lattice site from normalized Gaussian distributions, truncated at the 
values $0$ and $1$, with fixed mean (in most cases 
$\bar \mu = \bar \sigma = \bar \lambda = 0.5$) but different standard 
deviations $\sigma = 0 \ldots 0.9$.
The reaction rates therefore constitute quenched random variables.

\begin{figure}[!t]
\includegraphics[angle=0,width=3.3in]{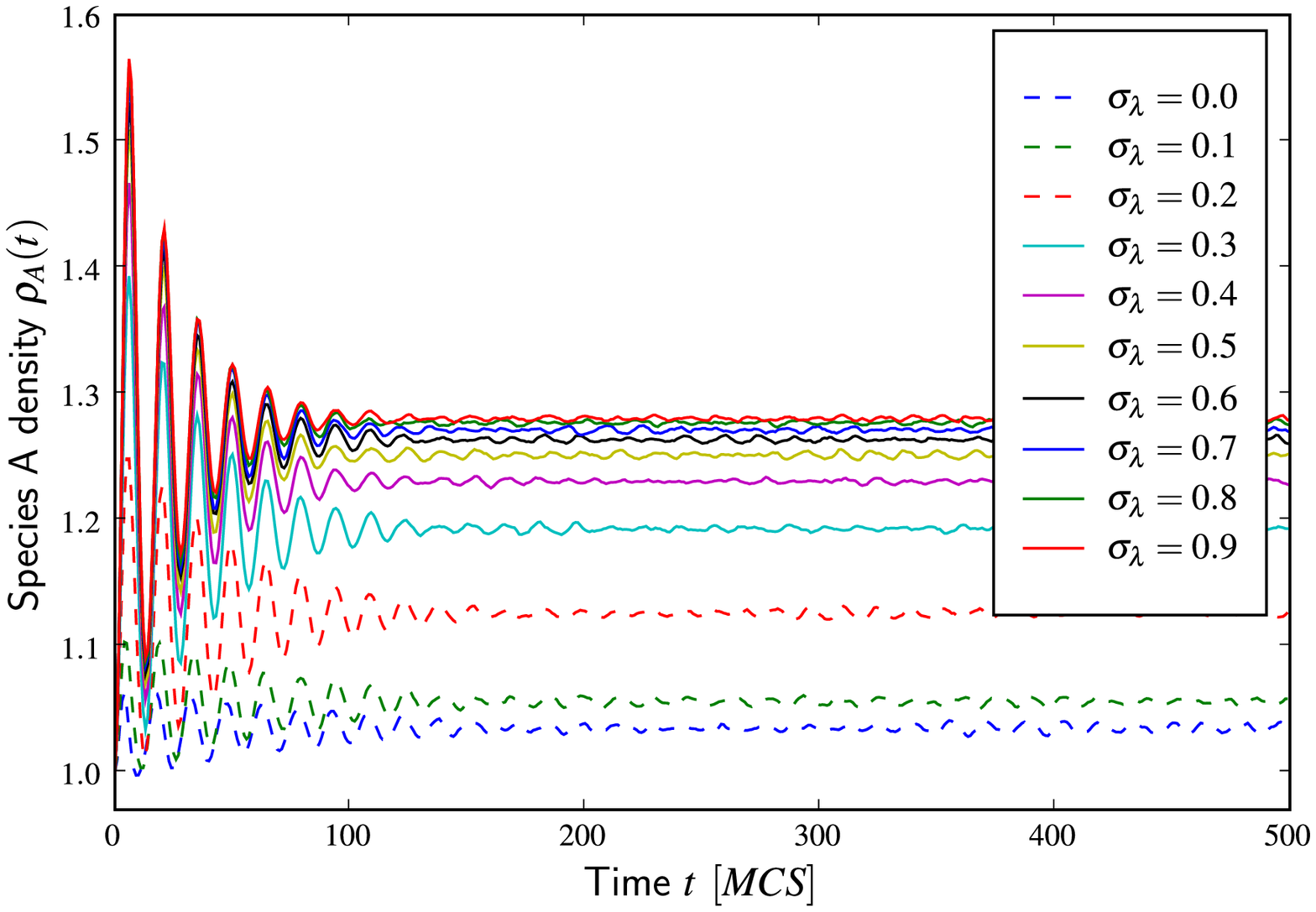} \vskip -0.1truecm
\includegraphics[angle=0,width=3.3in]{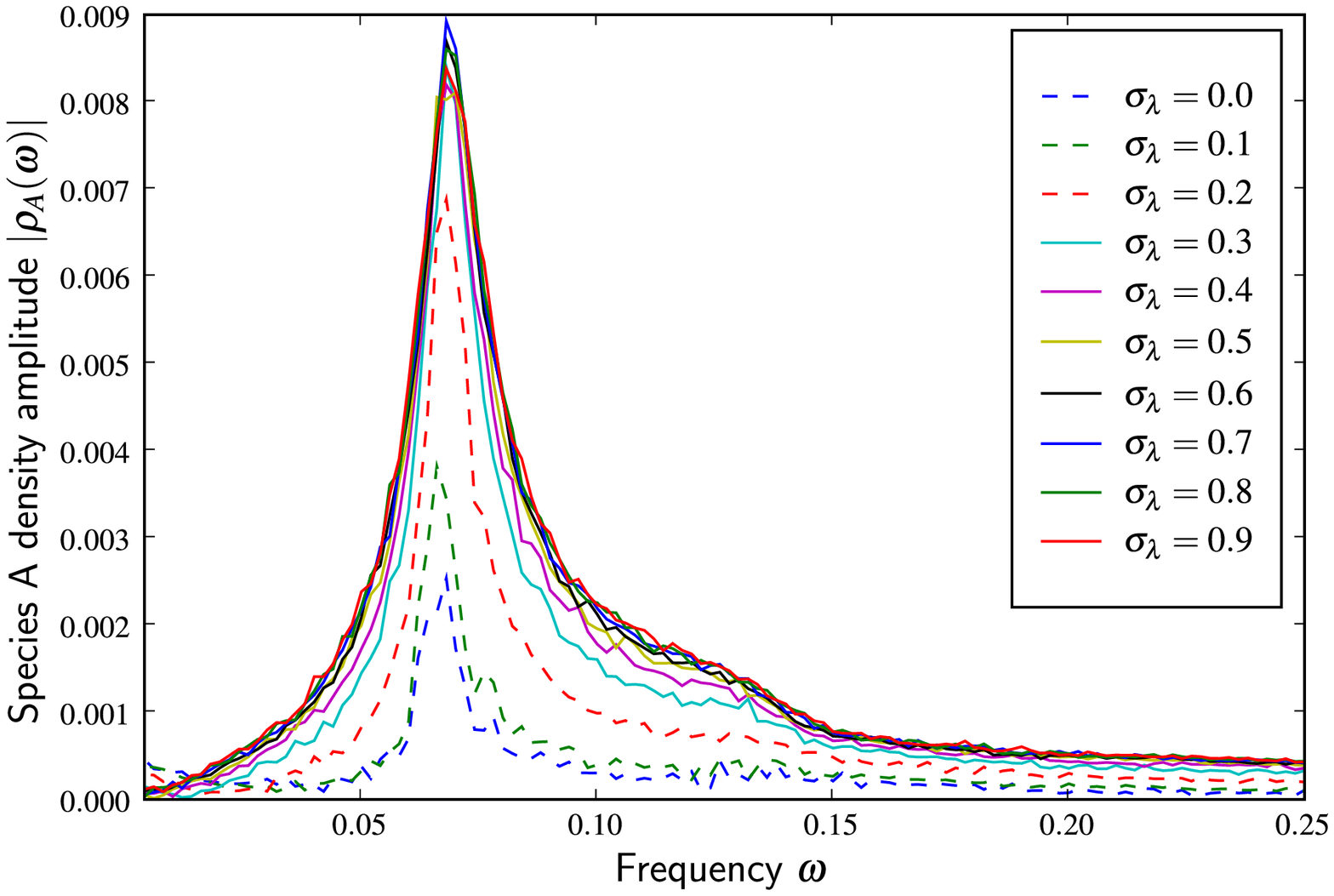} \vskip -0.2truecm
\caption{(a) Time evolution of the predator density $\rho_A(t)$, averaged over 
  $50$ Monte Carlo simulation runs on a 512 $\times$ 512 square lattice with 
  initial densities $\rho_A(0) = \rho_B(0) = 1$, predator death rate 
  $\mu = 0.5$, prey birth rate $\sigma = 0.5$, and mean predation rate 
  $\bar \lambda = 0.5$, for different variances $\sigma_\lambda$ as indicated. 
  (b) Signal Fourier transform $|\rho_A(\omega)|$. (Color online.)}
\label{pddens} \vskip -0.2truecm
\end{figure}
\begin{figure}[!t]
\includegraphics[angle=0,width=3.4in]{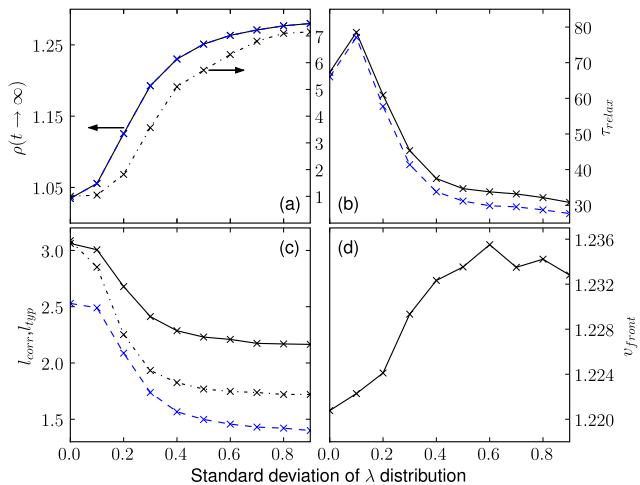} \vskip -0.1truecm
\caption{Dependence on the variance $\sigma_\lambda$, measured for
  uniform rates $\mu = \sigma = 0.5$ of (a) the asymptotic mean
  population densities $\rho_{A/B}(t \to \infty)$, compared with the
  average over the mean-field values (dashed-dotted lines, right-hand
  y-axis), (b) the relaxation time $\tau_{{\rm relax}\,A/B}$ towards
  the stationary state, (c) the predator/prey correlation lengths
  $l_{{\rm corr}\,A/B}$ -- predators $A$: full lines, prey $B$: dashed
  -- and the typical species separation distance $l_{\rm typ}$:
  dashed-dotted line; (d) the front speed $v_{\rm front}$ of the
  activity rings, obtained for $\mu = 0.2$ and $\sigma = 1.0$.}
\label{devcor} \vskip -0.2truecm
\end{figure}

The time evolution of the mean predator density 
$\rho_A(t) = \langle n_{A \, i}(t) \rangle$, averaged over $50$ Monte Carlo 
simulation runs with initially randomly placed particles with densities 
$\rho_A(0) = \rho_B(0) = 1$ is shown in Fig.~\ref{pddens}(a).
The absolute value of the Fourier transform (taken over the full interval of 
$500$ Monte Carlo steps) of this averaged signal, $|\rho_A(\omega)|$, is 
displayed in Fig.~\ref{pddens}(b).
Here we have used uniform rates $\mu = \sigma = 0.5$, while the predation rate
represents a quenched random variable with mean $\bar \lambda = 0.5$ and 
standard deviation $\sigma_\lambda$ ranging from $0$ to $0.9$ \cite{MovieFile}.
For these rates, the prey population density (not shown) behaves similarly, 
with an overall phase shift in the transient oscillations \cite{TransOsci}, and
both densities reach practically identical asymptotic density values, see also 
Fig.~\ref{devcor}(a).
It is evident that increasing spatial variability markedly amplifies the 
initial population oscillations and reduces the relaxation time towards the 
steady state.
Remarkably, {\em both} predator and prey densities approach larger asymptotic
values as $\sigma_\lambda$ is raised.
As shown in Fig.~\ref{devcor}(a), either species gains a remarkable fitness 
enhancement by $\sim 25 \%$ in the investigated $\sigma_\lambda$ range.
We have also studied spatial variations in the predator death rate $\mu$ and 
the prey birth rate $\sigma$, with the other rates held uniform.
In either case we observe merely a minute increase in the few percent range of 
the asymptotic predator and prey densities, not nearly as pronounced as the 
effect of spatially varying predation rates.

The neutrally stable species coexistence fixed point of the classical
Lotka--Volterra mean-field rate equations gives the stationary predator and 
prey densities as $\rho_A = \sigma  / \lambda$ and $\rho_B = \mu / \lambda$.
Presumably therefore, the fitness enhancement of both species stems from those
regions where the predation rate is significantly lower than the average.
Before we explore local effects in more detail, let us first consider a global 
average over the truncated Gaussian predation rate distributions of these 
mean-field stationary densities.
The result is depicted in Fig.~\ref{devcor}(a) along with the simulation data:
The `naive' averaging procedure indeed yields an increase of both stationary 
population densities; however, it predicts a grossly exaggerated fitness 
enhancement owing to the fact that mean-field approximations tend to 
overestimate cooperative effects \cite{CutoffEff}.
We therefore proceed to investigate the prominent role of spatial variations 
and predator--prey correlations in the lattice system.


As one would expect, increasing disorder broadens the peak associated with the 
transient oscillations in the associated Fourier signal, reflecting faster 
relaxation towards the asymptotic nonequilibrium stationary state.
Figure~\ref{pddens}(b) clearly reveals the roughly threefold increase in 
amplitude of the stochastic nonlinear population oscillations as 
$\sigma_\lambda$ is raised from $0$ to $0.9$.
By fitting the peak envelopes to a Lorentzian shape (which works well except in
the pure case with $\sigma_\lambda = 0$), we extracted the characteristic 
relaxation times $\tau_{{\rm relax}\,A/B} = 1 / \Gamma_{A/B}$ from the full 
widths at half maximum $\Gamma_{A/B}$ as function of $\sigma_\lambda$, see 
Fig.~\ref{devcor}(b). 
Note the reduction by a factor $\sim 2.5$ in $\tau_{{\rm relax}\,A/B}$ as 
$\sigma_\lambda$ is increased from $0.1$ to $0.6$.


The increasing amplitudes of the initial population oscillations suggest that
the spatial variability in the predation rates tends to cluster both species
closer together, thus enhancing localized population explosions.
This interpretation is in fact borne out by measuring the steady-state 
equal-time two-point correlation functions $C_{\alpha \beta}(x) = 
\langle n_{\alpha\, i+x} \, n_{\beta\, i} \rangle - \rho_\alpha \, \rho_\beta$
with $\alpha, \beta = A, B$ \cite{Corrfunct}.
After again averaging the data over $50$ Monte Carlo simulation runs, we have 
extracted the predator and prey correlation lengths $l_{{\rm corr}\,A/B}$,
which essentially measure the spatial extent of the population patches, as 
function of $\sigma_\lambda$ by least-square fits of $C_{AA}(x)$ and 
$C_{BB}(x)$ to exponentials $\exp(-|x| / l_{\rm corr})$ at sufficiently large 
distance $|x|$.
As depicted in Fig.~\ref{devcor}(c), the predator correlation length
$l_{{\rm corr}\,A}$ decreases by $\sim 30 \%$ from about $3$ to $2.1$ lattice 
constants as the disorder variance increases, while $l_{{\rm corr}\,B}$ is 
reduced by $\sim 45 \%$ from $2.5$ to $1.4$.

Similarly, we infer the typical predator--prey separation distance 
$l_{\rm typ}$ from the cross-correlation function $C_{AB}(x)$, which is 
negative at short distances, but attains a maximum with positive correlation
before tending towards $0$ as $|x| \to \infty$ \cite{Corrfunct}.
Here, we define $l_{\rm typ}$ as the location of the maximum of $C_{AB}(x)$.
The results as function of the standard deviation $\sigma_\lambda$, shown in
Fig.~\ref{devcor}(c), closely follow the behavior of the correlation lengths,
namely rather rapidly decreasing from $\sim 3$ lattice constants to $\sim 1.7$ 
as $\sigma_\lambda$ is reduced from $0.1$ to $0.4$.
Thus, when the width of the distribution of the spatially varying predation 
rates $\lambda$ becomes larger, the ensuing correlated patches of coexisting
predator--prey populations become more localized in regions with low values of
$\lambda$.
Consequently, a larger number of such patches can be accomodated in the system,
whereby the long-time population densities increase.
The stabilizing effect of spatial inhomogeneity has recently been elaborated in
a two-patch predator--prey model of diffusively coupled two-dimensional 
oscillators \cite{Yaari08}.


The classical two-species Lotka--Volterra reaction-diffusion equations, i.e., 
essentially the mean-field rate equations supplemented with diffusive 
spreading, are well-known to support traveling wave solutions 
\cite{Dunbar83, Sherratt97, Murray02, Aguiar04}, 
whose minimal front speed can be established by standard mathematical tools 
\cite{Hosono98, Mendez99, Lewis02}.
Beyond the mean-field approximation, however, already in single-species systems
the incorporation of intrinsic reaction noise in the computation of wave front
propagation velocities is a rather difficult problem 
\cite{Riordan95, Lemarchand95, Pechenik99, Panja04, Escudero04}, and there are 
very few results available for two-species models \cite{Panja04, Malley06}.

\begin{figure}[!t]
\includegraphics[angle=0,width=3.3in]{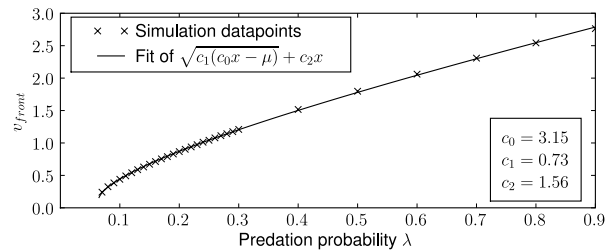} \vskip -0.1truecm
\caption{Propagation speed of radially spreading activity fronts in the 
  stochastic Lotka--Volterra model with uniform rates $\mu = 0.2$ and 
  $\sigma = 0$ as function of the predation rate $\lambda$. 
  The square-root fit is inspired by the mean--field lower bound.}
\label{frovel} \vskip -0.2truecm
\end{figure}
In the initial stage of our simulations, we observe radially spreading circular
fronts of prey followed by predators \cite{MovieFile}.
We have therefore set out to numerically measure the front propagation speed
of spreading rings of activity, namely prey invading empty regions followed by
predators feeding on them, in our two-dimensional stochastic Lotka--Volterra 
model.
To this end, we set up as initial state a circular patch of $B$ particles, one
per site, of radius $5$ lattice constants, and $10$ predators $A$ located on
the center site of this patch.
In this study, we have chosen uniform rates $\mu = 0.2$ and $\sigma = 1.0$,
with spatially varying predation rate with mean $\bar \lambda = 0.5$.
After angular averaging to obtain the radial particle concentrations, we have 
determined the invading front location by searching for the zero of the first 
derivative of the radial prey density.
A linear fit of the data with Monte Carlo time yields the front speed 
$v_{\rm front}$, which is then averaged over typically $50$ simulation runs. 
The change of propagation speed with the disorder variance $\sigma_\lambda$ is 
plotted in Fig.~\ref{devcor}(d).
We find a small but noticeable $\sim 1 \%$ increase of the spreading activity 
front speed as $\sigma_\lambda$ is raised from $0$ to $0.7$, which we interpret
as essentially a consequence of the larger amplitudes of the more localized 
population fluctuations caused by the spatial variability of the predation 
rate. 
Our results for spatially homogeneous rates are depicted as function of the
predation rate in Fig.~\ref{frovel}. 
To avoid problems at small $\lambda$ values due to prey population explosions, 
we chose as initial state a sea of unreproductive $B$ particles ($5$ per site)
and $5$ predators $A$ located on the center of the grid, with $\mu = 0.2$.
The data can be fitted reasonably well with a square-root expression
that is motivated by the known lower bound 
$v_{front} > \sqrt{4 D_A (\lambda' \rho - \mu)}$, where $\rho$ denotes the
prey carrying capacity \cite{Dunbar83, Sherratt97, Murray02}.
Here, $4 D_A = 1$, $\rho = 1000$, but the dimensionless reaction probability 
$\lambda' \rho \sim \lambda$, so indeed the fit constants $c_0$ and $c_1$ 
should be of order $1$, but capture fluctuation-induced renormalizations of the
mean-field parameters, and the additional offset $c_2 > 0$ describes the
deviation from the lower bound. 


In conclusion, we have employed Monte Carlo simulations to investigate a 
stochastic two-species Lotka--Volterra model subject to quenched disorder in 
the reaction rates on a two-dimensional lattice without occupation number 
restrictions.
While randomizing the prey birth and predator death rates has little effect, 
spatial variability in the predation / species invasion rate $\lambda$ markedly
enhances the asymptotic densities for both predator and prey populations.
We provide evidence that this remarkable fitness increase is caused by 
disorder-induced modifications in the emerging spatio-temporal structures:
Upon increasing the width of the random rate distribution, the typical length
scales of both the spatial predator--predator and the prey--prey correlations 
is reduced.
This results in more localized patches of activity, presumably in the vicinity
of regions where the local predation rates are smaller than their mean value.
The system is thus able to accomodate a larger amount of populated regions. 
We also find that spatial variability in the predation rate drastically 
amplifies the initial population oscillations and markedly reduces the time 
required to reach the steady-state configuration.
In contrast, the front speed of spreading activity rings from a localized 
center is not very strongly affected by the disorder.
Yet we do observe that the activity fronts accelerate slightly upon increasing
the variance of the predation rate.

This research has been supported in part through the IAESTE International
Student Exchange Program and the U.S. National Science Foundation, grant 
NSF-DMR 0308548.
We gratefully acknowledge inspiring discussions with J.~Banavar, R.~Kulkarni, 
M.~Mobilia, M.~Pleimling, B.~Schmittmann, N.~Shnerb, and R.~K.~P.~Zia.


\begin{thebibliography}{99}
\bibitem{Algorithm} 
  Additional details of the Monte Carlo simulation algorithm are described in 
  Ref.~\cite{Washenberger07}.
\bibitem{CutoffEff}
  We note that this mean-field average is sensitive to the cutoff at small 
  values of $\lambda$. This is obvious in the flat distribution limit 
  $\sigma_\lambda \to \infty$, which yields a logarithmic dependence on the 
  integration boundaries.
\bibitem{Corrfunct}
  For stochastic Lotka--Volterra models with spatially uniform rates, these
  correlation functions are depicted in Figs.~4--6 of Ref.~\cite{Mobilia07} and
  Fig.~5 in Ref.~\cite{Washenberger07}.
\end{thebibliography}
\end{document}